\documentclass[aps,pre,showpacs,groupedaddress]{revtex4}
\usepackage{amsfonts}
\usepackage{mathrsfs}
\usepackage{graphicx}
\usepackage{amsmath}
\usepackage{amssymb}
\usepackage{amsbsy}

\begin{document}

\title{Combined local search strategy for learning in networks of binary synapses}

\author{Haiping Huang$^{1}$}
\affiliation{$^1$Key Laboratory of Frontiers in Theoretical Physics
and $^2$Kavli Institute for Theoretical Physics China, Institute of
Theoretical Physics, Chinese Academy of Sciences, Beijing 100190,
China}
\author{Haijun Zhou$^{1,2}$}
\affiliation{$^1$Key Laboratory of Frontiers in Theoretical Physics
and $^2$Kavli Institute for Theoretical Physics China, Institute of
Theoretical Physics, Chinese Academy of Sciences, Beijing 100190,
China}
\date{\today}

\begin{abstract}
Learning in networks of binary synapses is known to be an
NP-complete problem. A combined stochastic local search strategy in
the synaptic weight space is constructed to further improve the
learning performance of a single random walker. We apply two
correlated random walkers guided by their Hamming distance and
associated energy costs (the number of unlearned patterns) to learn
a same large set of patterns. Each walker first learns a small part
of the whole pattern set (partially different for both walkers but
with the same amount of patterns) and then both walkers explore
their respective weight spaces cooperatively to find a solution to
classify the whole pattern set correctly. The desired solutions
locate at the common parts of weight spaces explored by these two
walkers. The efficiency of this combined strategy is supported by
our extensive numerical simulations and the typical Hamming distance
as well as energy cost is estimated by an annealed computation.

\end{abstract}

 \maketitle

The binary synaptic weight is more robust to noise and much simpler
for large-scale electronic implementations compared with its
continuous counterpart. However, learning in networks of binary
synapses is known to be an NP-complete problem~\cite{Blum-1992}.
This means that if one could find an algorithm to solve this problem
in a polynomial time, any other NP problems (a set of decision
problems that can be resolved in polynomial time on a
non-deterministic Turing machine) can also be solved in polynomial
time. For a single layered feed forward network with binary weights
(we call this network binary perceptron), the storage capacity was
predicted to be $\alpha_{s}\simeq0.83$~\cite{Krauth-1989} provided
that the number of weights $N$ tends to be infinity. Here we define
the ratio of the number of patterns $P$ to $N$ as the constraint
density $\alpha$ and the theoretical limit of  $\alpha$ for the
perceptron is termed the storage capacity. The binary perceptron
attempts to perform a random classification of $\alpha N$ random
input patterns~\cite{Engel-2001}. Many efforts have been devoted to
the nontrivial algorithmic issue of this difficult
problem~\cite{Kohler-1990,Patel-1993,Bouten-1998,Zecchina-2006,Baldassi-2007,Barrett-2008,Huang-2010jstat}.
For all finite $\alpha$, a discontinuous ergodicity breaking
transition for the binary perceptron at finite temperature is
predicted by the dynamic mean field theory~\cite{Horner-1992a}. When
the transition temperature is approached, the traditional simulated
annealing process is easily trapped by the suboptimal configurations
where a finite fraction of patterns are still not
learned~\cite{Patel-1993}. The difficulty of local search heuristics
for learning is likely to be connected to the fact that
exponentially many small clusters coexist in the weight space with a
more exponentially large number of suboptimal
configurations~\cite{Kaba-09}. Here, we define a connected component
of the weight space as a cluster of solutions in which any two
solutions are connected by a path of consecutive single-weight
flips~\cite{Arde-08}. A configuration of synaptic weight is
identified to be a solution if it is able to learn a prescribed set
of patterns. Various stochastic local search strategies by virtue of
random walks have been used to find solutions of constraint
satisfaction
problems~\cite{Semer-03,Bart-03,Alava-2008,Zhou-2009epjb,Krza-07,Lenka-2007pre}.
In our previous study~\cite{Huang-2010jstat}, we suggested a simple
sequential learning mechanism, namely synaptic weight space random
walking for the perceptronal learning problem. In this setting,
$\alpha N$ patterns are presented in a randomly permuted sequential
order and random walk of single- or double-weight flips is performed
until each newly added pattern is correctly classified (learned).
The previously learned patterns are \emph{not} allowed to be
misclassified in later stages of the learning process. This simple
sequential learning rule was shown to have good performances on
networks of $N\sim10^{3}$ synapses or less. The mean achieved
$\alpha$ is $0.57$ for $N=201$ and $0.41$ for $N=1001$.

\begin{center}
\begin{figure}
          \includegraphics[bb=112 555 477 748,width=8.0cm]{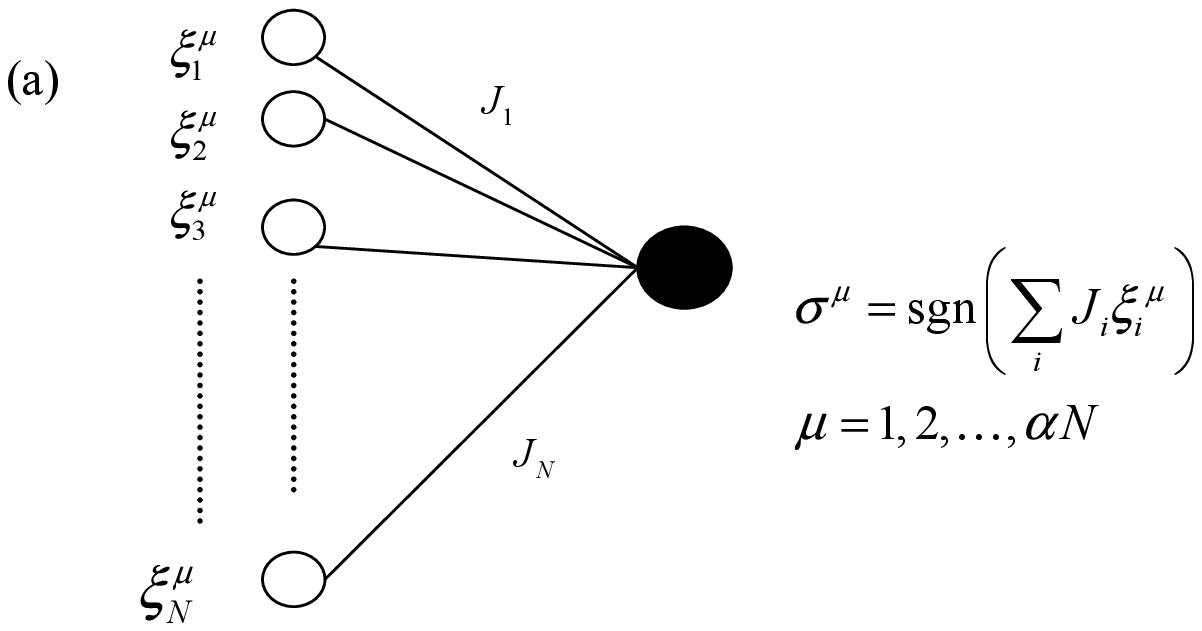}
     \hskip .5cm
     \includegraphics[bb=90 599 298 761,width=7.0cm]{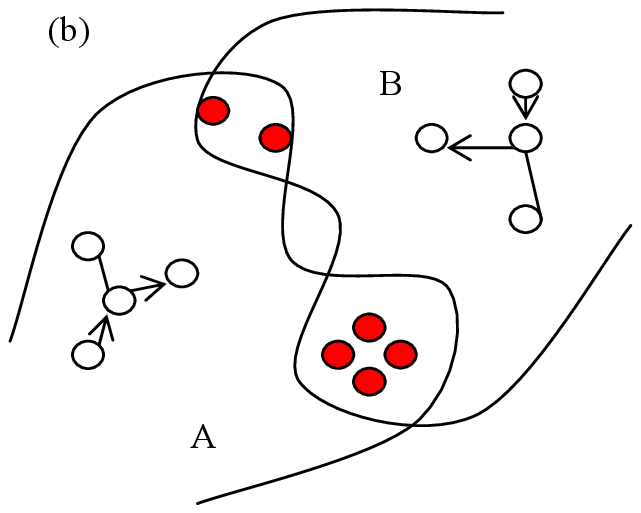}\vskip .2cm
  \caption{(Color online)
     The sketch of the binary perceptron and the multiple random walkings in the
     weight spaces. (a) $N$ input units (open circles)
     are connected directly to a single output unit (solid circle). A binary input pattern
    $(\xi_1^\mu, \xi_2^\mu, \ldots, \xi_N^\mu)$ of length $N$
    is transferred through a sign function to a binary output $\sigma^{\mu}$, i.e.,
    $\sigma^\mu = {\rm sgn}\bigl(\sum_{i=1}^{N} J_{i}\xi_{i}^{\mu}\bigr)$. The set of $N$
      binary synaptic weights $\{J_{i}\}$ is regarded as a solution of the perceptron
     problem if the output $\sigma^\mu= \sigma_0^\mu$ for each of the $P= \alpha N$ input patterns $\mu \in [1, P]$, where
     $\sigma_0^\mu$ is a preset binary value.
      (b) Mechanism of multiple random walkings. All configurations (represented by open or solid circles) are the
      solutions learning the initial small set of patterns for each
      random walker (A or B). The arrow indicates the movement for each
      walker
      in the weight space via SWF. The movement favors small Hamming distance between A and B and decreasing energy cost.
       The desired solutions (represented by solid circles) for the
      expected larger $\alpha$ locate at the common parts of weight
      spaces both walkers explore. Here two common parts are shown, but actually one or more than two are possible.
   }\label{perc}
 \end{figure}
 \end{center}

In this work, we improve the learning performance by introducing a
smart combined strategy. Instead of using a single random walker, we
apply two correlated random walkers guided by their Hamming distance
and associated energy costs. Both walkers expect to learn a same
large set of patterns, but each walker first learns a small part of
the whole pattern set (partially different for both walkers but with
the same constraint density). Then both walkers explore their
respective current weight spaces cooperatively until either of them
finds a solution to classify correctly the whole pattern set.
Therefore, each walker only needs to make the corresponding residual
part of the whole pattern set learned. Notice that the weight spaces
both walkers explore separately are actually different since the
small set of patterns they have learned are not completely identical
despite the same amount of learned patterns. If a solution exists,
the found solution should belong to one of the common parts of the
weight spaces explored by both walkers, therefore, the small Hamming
distance between these two walkers and zero energy are favored
during the multiple random walkings. In fact, the common parts will
appear as independent clusters once the larger expected $\alpha$ is
finally reached (see Fig.~\ref{perc}(b)).

The binary perceptron realizes a random classification of $P$ random
input patterns (see Fig.~\ref{perc}(a)). To be more precise, the
learning task is to seek for an optimal set of binary synaptic
weights $\{J_{i}\}_{i=1}^{N}$ that could map correctly each of
random input patterns $\{\xi_{i}^{\mu}\}(\mu=1,\ldots,P)$ to the
desired output $\sigma_{0}^{\mu}$ assigned a value $\pm1$ at random.
Given the input pattern $\boldsymbol{\xi}^{\mu}$, the actual output
$\sigma^{\mu}$ of the perceptron is $\sigma^{\mu}={\rm
sgn}\left(\sum_{i}J_{i}\xi_{i}^{\mu}\right)$ where $J_{i}$ takes
$\pm1$ and $\xi_{i}^{\mu}$ takes $\pm1$ with equal probability.  If
$\sigma^{\mu}=\sigma_{0}^{\mu}$, we say that the synaptic weight
vector $\boldsymbol{J}$ has learned the $\mu$-th pattern. Therefore
we define the number of patterns mapped incorrectly as the energy
cost
$E(\boldsymbol{J})=\sum_{\mu}\Theta\left(-\sigma_{0}^{\mu}\sum_{i}J_{i}\xi_{i}^{\mu}\right)$
where $\Theta(x)$ is a step function with the convention that
$\Theta(x)=0$ if $x\leq0$ and $\Theta(x)=1$ otherwise. In the
current setting, both $\{\xi_{i}^{\mu}\}$ and the desired output
$\{\sigma_{0}^{\mu}\}$ are generated randomly independently. Without
loss of generality, we assume $\sigma_{0}^{\mu}=+1$ for any input
pattern in the remaining part of this letter, since one can perform
a gauge transformation
$\xi_{i}^{\mu}\rightarrow\xi_{i}^{\mu}\sigma_{0}^{\mu}$ to each
input pattern without affecting the result.

Before introducing the combined learning strategy, we first briefly
outline two simple local search strategies~\cite{Huang-2010jstat},
i.e., single-weight flip (SWF) and double-weight flip (DWF). To
learn a given set of random patterns, we first generate an initial
weight configuration $(J_{1}^{0},J_{2}^{0},\ldots,J_{N}^{0})$ at
time $t=0$. The first pattern $\boldsymbol{\xi}^{1}$ is then
presented to the perceptron. If this pattern is correctly learned by
the initial weight configuration, then the second pattern
$\boldsymbol{\xi}^{2}$ is presented, otherwise the weight
configuration is modified by a sequence of SWF or DWF until
$\boldsymbol{\xi}^{1}$ is correctly classified. All patterns are
applied in a sequential order. Suppose at time $t$ the weight
configuration is
$\boldsymbol{J}^{t}=(J_{1}^{t},J_{2}^{t},\ldots,J_{N}^{t})$, and
suppose this weight configuration correctly classifies the first $m$
input patterns $\boldsymbol{\xi}^{\mu}(\mu=1,\ldots,m)$ but not the
$(m+1)$-th pattern $\boldsymbol{\xi}^{m+1}$. The random walker will
keep wandering in the weight space of the first $m$ patterns via SWF
or DWF until a configuration that correctly classifies
$\boldsymbol{\xi}^{m+1}$ is reached. In the SWF protocol, a set
$A(t)$ of allowed single-weight flips is constructed  based on the
current configuration $\boldsymbol{J}^{t}$ and the $m$ learned
patterns. $A(t)$ contains all integer indexes $j\in[1,N]$ with the
property that the single weight flip $J_{j}^{t}\rightarrow
-J_{j}^{t}$ does not make any barely learned patterns $\mu\in[1,m]$
(whose stability field $h^{\mu}=\sum_{i}J_{i}\xi_{i}^{\mu}=+1$)
being misclassified. At time $t'=t+1/N$, an index $j$ is chosen
uniformly randomly from set $A(t)$ and the weight configuration is
changed to $\boldsymbol{J}^{t'}$ such that $J_{i}^{t'}=J_{i}^{t}$ if
$i\neq j$ and $J_{j}^{t'}=-J_{j}^{t}$. The DWF protocol is very
similar to the SWF protocol with the only difference that the set
$A(t)$ contains pairs of integer indexes $(i,j)$ for allowed
double-weight flips. This set can be constructed as follows. For the
current configuration $\boldsymbol{J}^{t}$, if there are no barely
learned patterns ($h^{\mu}=+1$ or $+3$ for double-weight flips)
among the first $m$ learned patterns, $A(t)$ includes all the
$N(N-1)/2$ pairs of integers $(i,j)$ with $1\leq i<j\leq N$.
Otherwise, randomly choose a barely learned pattern, say
$m_{1}\in[1,m]$ and for each integer $i\in[1,N]$ with the property
$J_{i}^{t}\xi_{i}^{m_{1}}<0$, put $i$ into another set $B(t)$, then
do the following: (1) if $J_{i}^{t}\xi_{i}^{\mu}<0$ for all the
other barely learned patterns, then add all the pairs $(i,j)$ with
$j\not\in B(t)$ into the set $A(t)$; (2) otherwise, add all the
pairs $(i,j)$ into the set $A(t)$, with the property that the
integer $j\not\in B(t)$ satisfies $J_{j}^{t}\xi_{j}^{\mu}<0$ for all
those barely learned patterns $\mu\in[1,m]$ with
$J_{i}^{t}\xi_{i}^{\mu}>0$. In practice, when the number of added
patterns is small, we use an alternative scheme where the set $A(t)$
is not pre-constructed and instead we randomly select a pair $(i,j)$
and flip them if the flip would not make any previously learned
patterns misclassified. Once the number of flippable pairs of
weights is substantially reduced, we will use the DWF protocol
described above to keep learning proceeding. In this way, a learning
of a relatively large set of patterns would be not very time
consuming.

The combined strategy to improve the learning performance of single
walker is illustrated in Fig.~\ref{perc}(b). Before the learning
starts, we divide the given set of patterns to be learned into three
parts, namely $\mathcal {A},\mathcal {B}$ and $\mathcal {C}$ with
the property that the number of patterns in $\mathcal
{A}\cup\mathcal {B}$ equals to that in $\mathcal {A}\cup\mathcal
{C}$. Then the first walker tries to learn $\mathcal {A}\cup\mathcal
{B}$ by DWF protocol while the second walker tries to learn
$\mathcal {A}\cup\mathcal {C}$ by the same protocol. After $\mathcal
{A}\cup\mathcal {B}$ and $\mathcal {A}\cup\mathcal {C}$ have been
learned by both walkers separately, each walker keeps wandering in
its current weight space via SWF with the property that all
previously learned patterns are still learned and no new pattern is
added. In addition, they should communicate with each other by
lowering down the sum of the Hamming distance and the associated
energy costs. If the sum goes up, we accept the walking (both
walkers modify their current configurations via SWF one time) with
the probability $e^{-\beta N(\Delta H_{d}+\Delta e)}$ where $\Delta
H_{d}$ is the change of Hamming distance with respect to the walking
and $\Delta e$ the change of energy cost density. $\beta$ serves as
a control parameter to be optimized. One could also introduce
another inverse temperature $\gamma$ (see Eq.~(\ref{Z})) to control
the decreasing rate of energy
cost~\cite{Hopfield-1982,Willshaw-1969,Willshaw-1991}. For
simplicity, we set $\gamma=\beta$ in our simulations although they
can be changed independently.
$H_{d}=\frac{1}{2}\left(1-\frac{1}{N}\sum_{i}J_{i}^{(1)}J_{i}^{(2)}\right)$
where $\boldsymbol{J}^{(1)}$ is the current weight configuration of
the first walker while $\boldsymbol{J}^{(2)}$ the second walker. For
the first walker, the energy cost is the number of patterns in
$\mathcal {C}$ misclassified by $\boldsymbol{J}^{(1)}$ and the
energy cost for the second walker is the number of patterns in
$\mathcal {B}$ misclassified by $\boldsymbol{J}^{(2)}$. Once either
of these two energy costs becomes zero, the whole learning process
will be terminated and the whole set of patterns is learned.
Otherwise, the learning process stops if the maximal number of
attempts for multiple random walkings, namely $\mathcal {T}_{max}$,
is saturated. $\mathcal {T}_{max}$ is a free parameter whose value
should be chosen considering the trade-off between efficiency (a
solution is found) and computational cost.

The combined local search strategy through multiple random walkings
actually utilizes the smoothness of the weight space at the initial
low constraint density to achieve the solution at expected high
constraint density. In this process, the Hamming distance between
both walkers and their associated energy costs play a key role in
guiding either (or both) of them to the desired solution. To get a
preliminary estimate of the optimal inverse temperature $\beta$, we
perform an annealed computation which is able to give us a crude
knowledge of the relation between the Hamming distance (or energy
cost) and the inverse temperature. In our current setting, we can
write the partition function as:
\begin{equation}\label{Z}
\begin{split}
Z&=e^{-\frac{1}{2}N\beta}\sum_{\boldsymbol{J}^{(1)},\boldsymbol{J}^{(2)}}e^{\frac{\beta}{2}\sum_{i}J_{i}^{(1)}J_{i}^{(2)}}
\prod_{\mu\in\mathcal {A}\cup\mathcal
{B}}\Theta\left(\frac{1}{\sqrt{N}}\sum_{i}J_{i}^{(1)}\xi_{i}^{\mu}\right)\\
&\cdot\prod_{\mu\in\mathcal
{C}}\left[e^{-\gamma}+(1-e^{-\gamma})\Theta\left(\frac{1}{\sqrt{N}}\sum_{i}J_{i}^{(1)}\xi_{i}^{\mu}\right)\right]\\
&\cdot
\prod_{\mu\in\mathcal {A}\cup\mathcal {C}}\Theta\left(\frac{1}{\sqrt{N}}\sum_{i}J_{i}^{(2)}\xi_{i}^{\mu}\right)\\
&\cdot\prod_{\mu\in\mathcal
{B}}\left[e^{-\gamma}+(1-e^{-\gamma})\Theta\left(\frac{1}{\sqrt{N}}\sum_{i}J_{i}^{(2)}\xi_{i}^{\mu}\right)\right]
\end{split}
\end{equation}
where we still distinguish $\beta$ and $\gamma$ which are set to be
equal in our simulations. Note that the added prefactor $N^{-1/2}$
makes the argument of $\Theta(\cdot)$ of order of unity for the sake
of statistical mechanics analysis. In annealed approximation, we
compute the disorder average of partition function $\left<Z\right>$
where$\left<\cdots\right>$ denotes the average over the input random
patterns. We skip the detail of computation here and define the
overlap between configurations $\boldsymbol{J}^{(1)}$ and
$\boldsymbol{J}^{(2)}$ as
$q=\frac{1}{N}\sum_{i}J_{i}^{(1)}J_{i}^{(2)}$, then the annealed
approximated free energy density $f_{ann}$ is given by:
\begin{equation}\label{Z2}
\begin{split}
-\beta f_{ann}&=\frac{\log\left<Z\right>}{N}=\max_{q,\hat{q}}\Biggl\{-\frac{\beta}{2}-q\hat{q}+\frac{\beta}{2}q+\log(4\cosh\hat{q})\\
&+\alpha_{c}\log\Bigl[\int_{0}^{\infty}Dt H\bigl(-\frac{qt}{\sqrt{1-q^{2}}}\bigr)\Bigr]\\
&+(\alpha-\alpha_{c})\log\Bigl[e^{-\gamma}H(0)+(1-e^{-\gamma})\\
&\cdot\int_{0}^{\infty}Dt H(-\frac{qt}{\sqrt{1-q^{2}}})\Bigr]
\Biggr\}
\end{split}
\end{equation}
where $H(x)=\int_{x}^{\infty}Dt$ and
$Dt\equiv\frac{dt}{\sqrt{2\pi}}e^{-t^{2}/2}$. $\alpha_{c}$ is the
common constraint density denoted by the ratio of the number of
patterns in the set $\mathcal {A}$ to $N$, and the set $\mathcal
{A}$ is the common part learned by both walkers at the initial
stage. $\hat{q}$ is the conjugate counterpart of overlap $q$ and
both of them are determined by the following recursive equations:
\begin{subequations}\label{q}
\begin{align}
q&=\tanh\hat{q}\\
\begin{split}
\hat{q}&=\frac{\beta}{2}+\frac{\alpha_{c}}{\sqrt{1-q^{2}}{\rm
arccot}\left(-\frac{q}{\sqrt{1-q^{2}}}\right)}\\
&+\frac{(\alpha-\alpha_{c})(1-e^{-\gamma})}{\sqrt{1-q^{2}}\left[\pi
e^{-\gamma}+(1-e^{-\gamma}){\rm
arccot}\left(-\frac{q}{\sqrt{1-q^{2}}}\right)\right]}
\end{split}
\end{align}
\end{subequations}
After the solution of the above recursive equations is obtained, the
annealed typical Hamming distance is calculated via
$H_{d}^{ann}=\frac{1-q}{2}$ and the annealed energy density is
evaluated as:
\begin{equation}\label{e}
    e_{ann}=\frac{(\alpha-\alpha_{c})\left[\pi-{\rm
arccot}\left(-\frac{q}{\sqrt{1-q^{2}}}\right)\right]}{\pi-{\rm
arccot}\left(-\frac{q}{\sqrt{1-q^{2}}}\right)+e^{\gamma}{\rm
arccot}\left(-\frac{q}{\sqrt{1-q^{2}}}\right)}
\end{equation}
In practical learning of a single instance, we choose the optimal
temperature where predicted $H_{d}^{ann}$ and $e_{ann}$ take small
values (e.g., around $0.06$ and $0.01$ respectively) and these
predicted values can also be compared with those obtained during the
actual learning processes. Note that the learning performance is not
very sensitive to small changes in the temperature as long as the
used temperature yields relatively small predicted $H_{d}^{ann}$ and
$e_{ann}$ . In addition, we define an initial constraint density
$\alpha_{I}$ as the number of patterns in $\mathcal
{A}\cup\mathcal{B}$ or $\mathcal {A}\cup\mathcal{C}$ over $N$. Its
value should be chosen to be relatively small such that each walker
can learn the initial pattern set and the common parts of weight
spaces both walkers explore exist. In practice, we choose
$\alpha_{I}=0.4,0.35,0.3$ for $N=201,501,1001$ respectively.

\begin{figure}
          \includegraphics[bb=14 12 249 138,width=8.0cm]{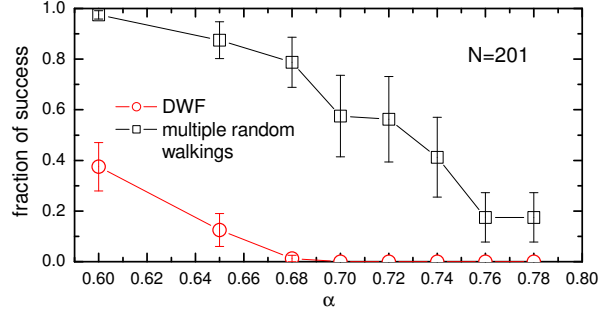}
  \caption{(Color online)
        Comparison of learning performances using DWF and multiple
        random walkings. The pattern length is $N=201$. For each
        sample, we try to learn a fixed set of random unbiased
        patterns ten times and record the fraction of success. The
        error bar indicates the fluctuation across eight random
        samples. We choose the optimal temperature according to
        Eq.~(\ref{e}) with relatively small predicted $H_{d}^{ann}$
        and $e_{ann}$. We set $\mathcal {T}_{max}=5\times10^{4}N$,
        and the initial constraint density
        $\alpha_{I}=0.4$.
   }\label{comp}
 \end{figure}

\begin{figure}
          \includegraphics[bb=16 19 265 140,width=8.0cm]{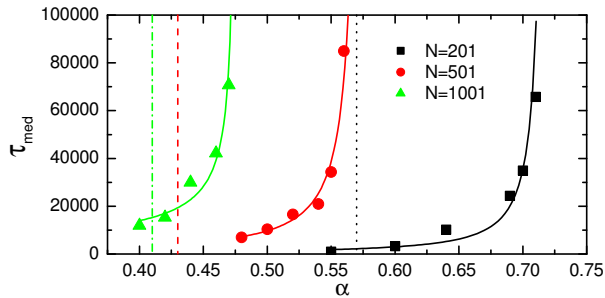}
  \caption{(Color online)
  Median learning time $\tau_{med}$ versus constraint density $\alpha$ using
  multiple random walkings for different $N$. $16$ random pattern sets are generated and the corresponding
  learning times (walking steps) are ordered; $\tau_{med}$ is the median of this ordered sequence. Those cases where the learning fails within $\mathcal {T}_{max}$
  are put at the top of the ordered learning time sequence. $\mathcal {T}_{max}=5\times10^{4}N,2\times10^{4}N,10^{4}N$ for
  $N=201,501,1001$ respectively. We choose the optimal temperature according to
  Eq.~(\ref{e}) with relatively small predicted $H_{d}^{ann}$
  and $e_{ann}$. The solid lines are power-law
  fittings of the form
  $\tau_{med}\propto(\alpha_{cr}-\alpha)^{-\delta}$ where $\alpha_{cr}\simeq0.72,0.575,0.475$, and $\delta\simeq1.374,1.257,0.644$ for $N=201,501,1001$ respectively.
  The dashed-dotted line indicates the mean
  constraint density achieved by DWF for $N=1001$, dashed line for $N=501$ and dotted line for $N=201$~\cite{Huang-2010jstat}.
   }\label{medtime}
 \end{figure}

\begin{figure}
          \includegraphics[bb=12 13 245 141,width=7.5cm]{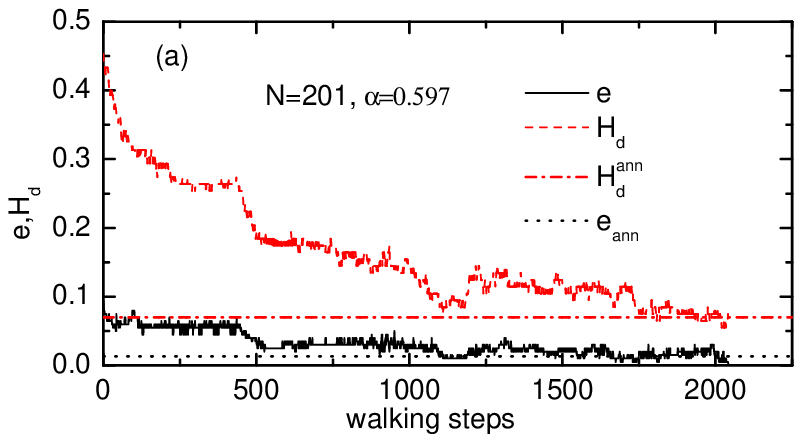}
     \vskip .2cm
     \includegraphics[bb=13 15 266 137,width=7.5cm]{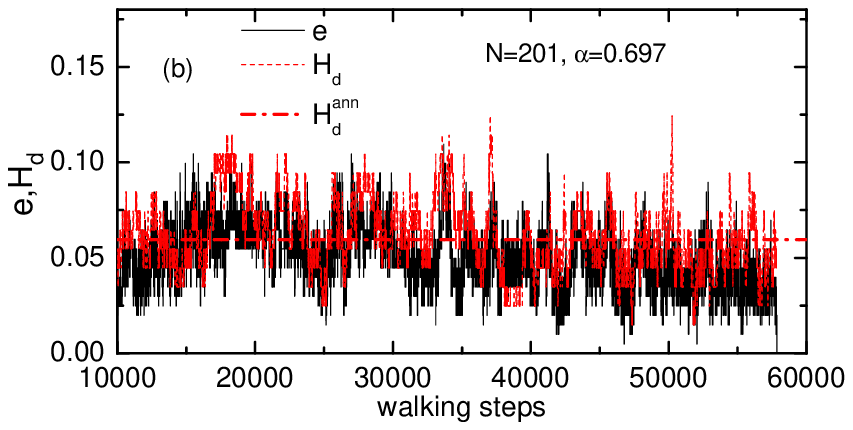}\vskip .2cm
     \includegraphics[bb=13 15 266 137,width=7.5cm]{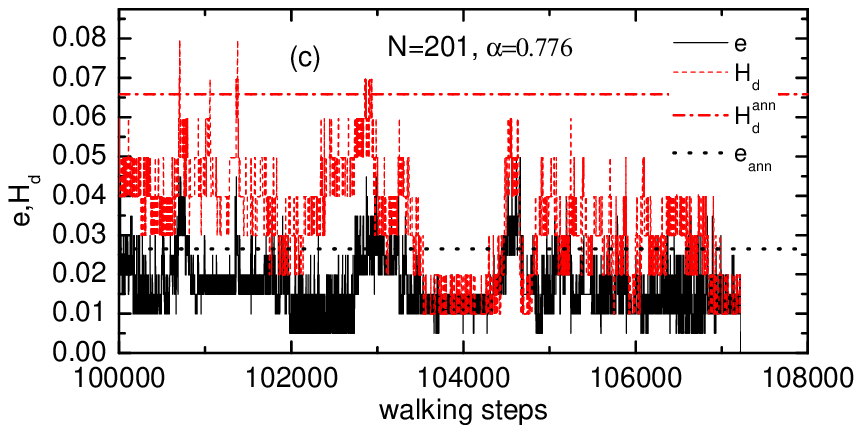}\vskip .2cm
     \includegraphics[bb=13 12 265 138,width=7.5cm]{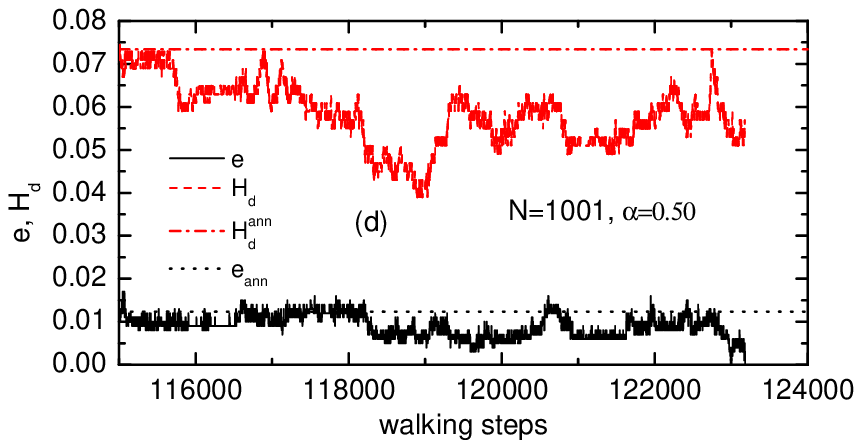}\vskip .2cm
  \caption{(Color online)
     Evolution of Hamming distance and energy cost for multiple
     random walkings to find a solution of single instance at high constraint density.
     The evolution corresponds to the walker finding the desired solution. (a) $N=201,\alpha=0.597$. The dashed-dotted
     line stays for the annealed Hamming distance while the dotted line the
     annealed energy cost. The temperature is chosen according to
     Eq.~(\ref{e}) with relatively small predicted $H_{d}^{ann}$
     and $e_{ann}$. $\alpha_{I}=0.4, \beta=1.8,\mathcal
     {T}_{max}=5\times10^{4}N$. (b) The same as (a), but both walkers are guided only by Hamming
     distance and the temperature is chosen with relatively small predicted
     $H_{d}^{ann}$. $\alpha=0.697,\beta=2.6$. (c) The same as (a), but for
     $\alpha=0.776,\beta=1.7$. (d) $N=1001,\alpha=0.50,\alpha_{I}=0.3,\beta=1.9,\mathcal
     {T}_{max}=5\times10^{3}N$.
   }\label{Evo}
 \end{figure}

We apply the proposed combined local search strategy to learn random
patterns of length $N=201$ and compare the result with that obtained
by DWF. DWF is able to go on even if SWF does not work, i.e., all
weights are frozen but flipping certain pairs of weights is still
permitted. Hence DWF can achieve higher mean $\alpha$ than SWF.
Furthermore, its learning time grows almost linearly with $\alpha$
up to the constraint density where DWF can not proceed any
more~\cite{Huang-2010jstat}. Therefore we only consider the
comparison between combined local search strategy and DWF. As shown
in Fig.~\ref{comp}, multiple random walkings do outperform DWF
despite the large observed fluctuation across different samples at
large expected $\alpha$. In fact, DWF is easily trapped if the
current configuration is frozen with respect to double-weight flips,
which occurs with high probability when the constraint density
becomes large~\cite{Huang-2010jstat}. It is expected that the weight
space will become rather rugged at high $\alpha$ in the sense that
exponentially many small solution clusters appear and most of them
can not be connected by simple single-weight or double-weight flip.
However, multiple random walkings start to find the solution for a
high $\alpha$ from the weight space at a relatively small
$\alpha_{I}$ where a big connected component each walker will probe
is expected. To achieve the desired solution in available time
scales, the Hamming distance between both walkers and the associated
energy costs are needed to guide both walkers, since the solutions
for the high $\alpha$ locate at the common parts of weight spaces
both walkers explore and these common parts will emerge as
independent clusters when the high $\alpha$ is finally reached. As
shown in Fig.~\ref{comp}, the combined strategy still has a finite
probability to learn $0.78N$ input random patterns while DWF is not
able to learn patterns with $\alpha\geq0.70$. Even at small
constraint density $\alpha\simeq0.60$, the fraction of learning
success by multiple random walkings can be nearly $100\%$ with less
fluctuations and very small walking steps (see Fig.~\ref{Evo} (a)).
In Fig.~\ref{medtime}, we also show the median learning time
(walking steps) of $16$ random pattern sets. This is done by
recording walking steps needed to learn each pattern set, then the
learning times are ordered~\cite{Priel-94}. We define $\tau_{med}$
as the median value of this ordered sequence. Those cases where the
learning fails within $\mathcal {T}_{max}$ are put at the top of the
ordered learning time sequence. As shown in Fig.~\ref{medtime}, the
critical constraint density $\alpha_{cr}$ at which $50\%$ of the
presented pattern sets are learned successfully is larger than the
mean one achieved by DWF~\cite{Huang-2010jstat}.
$\alpha_{cr}\simeq0.72,0.575,0.475$ for $N=201,501,1001$
respectively, and $\tau_{med}$ grows with $\alpha$ roughly as a
power law $\tau_{med}\propto(\alpha_{cr}-\alpha)^{-\delta}$. As $N$
increases, the critical value $\alpha_{cr}$ decreases, which is
consistent with the fact that the quality of a polynomial algorithm
for the binary perceptron decreases with increasing system
size~\cite{Horner-1992a,Patel-1993}. However, the combined local
search strategy does improve the learning performance of a single
random walker.

Fig.~\ref{Evo} gives the evolution of Hamming distance and energy
cost for multiple random walkings to find a solution of a single
instance at high constraint density. For $N=201$, one can see from
Fig.~\ref{Evo}(a) that a solution for $\alpha\simeq0.597$ can be
found within $2000$ walking steps. In addition, the annealed
computation reproduces the plateau values of Hamming distance and
energy cost with very good agreement. If both walkers are guided
only by Hamming distance, the solution can also be found for
$\alpha\simeq0.70$ and $N=201$, but the fraction of success is
reduced to $33.8\%\pm16.5\%$ with $\beta=2.6$. As displayed in
Fig.~\ref{Evo}(b), the evolution of $H_{d}$ seems to be highly
correlated (almost synchronous) with that of the energy cost and
$H_{d}^{ann}$ is consistent with the plateau value of Hamming
distance. Without the guide of Hamming distance, both walkers are
easily trapped by suboptimal configurations with a small finite
energy. However, guided by both Hamming distance and energy,
searching for a solution can be speeded up since the absolute value
of $N\Delta H_{d}$ is usually comparable to that of $N\Delta e$
during the learning process. Notice that most of weight
configurations in weight spaces explored by both walkers act as
suboptimal configurations for the learning problem at expected high
constraint density and there may exist very narrow corridors to the
common part where the desired solutions belong to. If we apply a
single walker to explore its weight space after the initial stage to
find a solution and this walker is guided only by energy, we found
it easily gets stuck in local minima of energy landscape as well.
Fig.~\ref{Evo} (c) plots the evolution of Hamming distance and
energy cost during the whole learning process for a very high
constraint density $\alpha\simeq0.776$ for which a larger number of
walking steps to achieve the desired solution are required. For
$N=1001$ and expected $\alpha=0.50$, it is very difficult to find a
solution through two correlated random walkers provided that the
maximal number of walking attempts is limited. However, we still
found a solution with $\mathcal {T}_{max}=5\times10^{3}N$ and the
evolution of $H_{d}$ and $e$ is presented in Fig.~\ref{Evo} (d). In
Fig.~\ref{Evo} (c) and (d), our annealed estimates of $H_{d}$ and
$e$ seem to be larger than the actual plateau values, however, they
still provide us the information to select the optimal control
parameter $\beta$. To more accurately predict the actual plateau
values of $H_{d}$ and $e$, a quenched computation within replica
symmetry approximation or one-step replica symmetry breaking
approximation is needed~\cite{Krauth-1989}, which we leave for
future work.

In conclusion, we apply two correlated random walkers instead of
single walker to improve learning performance of the binary
perceptron. The solution for small $\alpha_{I}$ can be easily
obtained by SWF or DWF~\cite{Huang-2010jstat}. Both walkers then
explore their respective weight spaces cooperatively to reach one of
the common parts to which the solution for high $\alpha$ belongs.
The smart combined strategy through multiple random walkings makes
the whole weight space at expected high constraint density ergodic
for both walkers and the learning in the most hard phase becomes
possible. The efficiency of our method depends on the choice of a
suitable temperature to help the walker overcome energy or entropic
barriers and also relies on the smoothness of the initial weight
space at $\alpha_{I}$ whose value should ensure the common parts of
weight spaces both walkers explore exist. To this end, we derive
annealed estimations (Eq.~(\ref{q}) and Eq.~(\ref{e})) to select the
suitable temperature with small predicted $H_{d}^{ann}$ and
$e_{ann}$. Interestingly, the Hamming distance is found to be
important for guiding correlated walkers to find a solution (see
Fig.~\ref{Evo} (b)). However, as $N$ (also expected $\alpha$)
increases (see Fig.~\ref{medtime}), the learning time to reach the
desired solution grows rapidly, or a much larger $\mathcal
{T}_{max}$ should be preset. This supports the computational
difficulty to find solutions for binary perceptron by virtue of
local search
heuristics~\cite{Horner-1992a,Patel-1993,Huang-2010jstat}. Future
research is needed to acquire a full understanding of this point.



\acknowledgments  We thank Chuang Wang and Jinhua Zhao for careful
readings of the manuscript. The present work was in part supported
by the National Science Foundation of China (Grant numbers 10774150
and 10834014) and the China 973-Program (Grant number 2007CB935903).



\end{document}